\documentclass[aip,graphicx]{revtex4-1}
\usepackage{graphicx}
\usepackage{amsmath}
\usepackage{subfigure}

\begin{document}


\title{Coupled Modal-Nonmodal Interactions Due to Periodic, Infinite Train of Convecting Vortices (TCV)} 




\author{Jyothi Kumar Puttam}
\affiliation{Computational \& Theoretical Fluid Dynamics, CSIR-NAL, Bangalore - 560017, India}
\affiliation{PhD Scholar, High Performance Computing Laboratory, Aerospace Engineering Dept., IIT Kanpur, UP - 208016, India}
\author{Prasannabalaji~Sundaram}
\affiliation{CERFACS, Toulouse, France}
\author{Vajjala K. Suman}
\affiliation{Computational \& Theoretical Fluid Dynamics, CSIR-NAL, Bangalore - 560017, India}
\author{Ankan~Sarkar}
\affiliation{Department of Mechanical Engineering, IIT (ISM) Dhanbad, Jharkhand - 826 004, India}
\author{Tapan~K.~Sengupta}
\affiliation{High Performance Computing Laboratory, IIT Kanpur, UP - 208016, India}
\author{Tirupathur~N.~Venkatesh}
\affiliation{Computational \& Theoretical Fluid Dynamics, CSIR-NAL, Bangalore - 560017, India}
\author{Rakesh~K.~Mathpal}
\affiliation{Non-equilibrium Flow Simulation Laboratory, IIT Kanpur, UP - 208016, India}


\date{\today}

\begin{abstract}
Events during transition to turbulence either follow modal or non-modal routes, or combinations of the two. Here, we report a computational investigation of strong freestream excitation caused by a train of convecting vortices. For this TCV excitation, we show a strong interaction of modal and non-modal components causing a spectacular growth of disturbances. We propose this as the mechanism for the severe encounters due to convective vortical disturbances on the underlying shear layer.
\end{abstract}


\maketitle 

\section{Introduction}
\label{intro}
To understand the encounters of convecting small scale vortical structures with underlying shear layers, one must understand the subtle differences between instability and receptivity, i.e. the response of a dynamical system to specific classes of inputs as described in the textbook \cite{Sengupta21}. This began with the classical pipe flow experiment of Reynolds \cite{Reynolds83}, who delayed the onset of turbulence significantly by controlling ambient parameters during the experiments. The search for the highest critical parameters (Reynolds number) in a pipe flow has yet to be established due to interdependence between instability and receptivity due to present levels of ambient disturbances. Similarly, all encounters between aircraft and convective vortices in the free stream turbulence depend upon relevant parameters. The purpose here is to advance our understanding of turbulence encounters (modelled as train of convecting vortices) as a growth of disturbances during receptivity and instability of fluid flow over a flat plate, which is modeled as the surface of the aircraft wing.

Historically, disturbance growth in viscous flows have been viewed to be either via modal route \cite{DrazinReid81,BetchovCrim67,SchmidHenn01,Chandrasekhar61} or via nonmodal route \cite{Chomaz05,Kerswell18,Schmid07,TrefethenTrefReddDris93}. The search of the modal route followed the solution of Orr-Sommerfeld equation (OSE) \cite{Orr07,Sommerfeld09} that led to the finding of Tollmien-Schlichting (TS) waves \cite{Tollmienwave,Schlichting33,R_Jordinson_JFM}, and it was presumed that the TS waves are the precursor of transition to turbulence. Most of the modal route experimental studies are related to finding the TS waves caused by wall excitation \cite{SchubauerSkra47}. However there are experimental and theoretical investigations in \cite{GasterGran75,Sundaram22,SundaramSengSeng19}, where the authors created transition without creating any TS waves by pulse excitation of a boundary layer. In \cite{GasterGran75}  it was presumed such nonmodal growth as combinations of TS modes which was corrected in \cite{Sundaram22,SundaramSengSeng19} to be strictly due to transition due to nonmodal growth as spatio-temporal wave front, even when the boundary layer was monochromatically excited by localized manner. Most of the nonmodal studies in recent times \cite{Chomaz05,Kerswell18,Schmid07,TrefethenTrefReddDris93} followed the classification in \cite{Nishioka_Morkovin} where the authors propounded bypass transition as the event that completely precludes TS waves. Other researchers have extensively studied bypass transition in  \cite{SchmidHenn01,Bypass_BrandtSchlatterHenningson,Bypass_DurbinWu,Bypass_JacobsDurbin,Bypass_SaricReedKerschen,Bypass_Zaki,Bypass_ZakiDurbin}.

In the literature \cite{DrazinReid81, SchmidHenn01} no distinction is generally made between wall and freestream excitation and it is presumed that the freestream excitation creates an equivalent wall excitation that gives rise to TS waves. However, even earlier researchers, as in \cite{Nishioka_Morkovin,Taylor36, MoninYagl71}, perceived flow transition strictly from the perspective of freestream excitation that causes unsteady pressure perturbation inside the shear layer. This was also referred to as bypass transition in \cite{Nishioka_Morkovin}. In recent times the freestream excitation of boundary layer problem has been shown 
\cite{SenguptaSundSeng20} the transition to turbulence to occur as a receptivity problem following nonmodal, nonlinear global route. 

In recent times the present authors have demonstrated the route to turbulence by wall excitation starting from the solution of OSE to full nonlinear compressible Navier-Stokes equations (CNSE), the simultaneous presence of modal and nonmodal components of disturbance growth as a spatio-temporal wave front (STWF) \cite{Sundaram22,Sengupta12,SenguptaBallNijh94,SenguptaRaoVenk06a}, unlike the previous approach of studying either spatial \cite{Mack84} or temporal route \cite{BetchovCrim67}. For the monochromatic wall excitation problem, it was noted that the nonmodal component (STWF) dominates over the TS wave \cite{SenguptaRaoVenk06a,SenguptaBhau11}. Sengupta and co-authors also studied receptivity of boundary layer to freestream convective excitation \cite{SenguptaSundSeng20, SenguptaSumaSeng19, Sengupta20}.
Experimental demonstration of freestream excitation was reported by Kendall \cite{Kendall87,Kendall90} for a freestream train of convecting vortices (TCV). Vortex-induced disturbance growth by a single vortex in the freestream was demonstrated by receptivity experiments in \cite{LimSengChat04}, which was numerically shown in \cite{SenguptaSumaSeng19, SenguptaDeSark03} by solving the full, nonlinear incompressible Navier-Stokes equation. The freestream excitation cases for vortex-induced disturbance growth showed nonmodal route only. The physical mechanism of transition by freestream excitation caused by a single convecting vortex was shown by solving OSE and linearized Navier-Stokes equation \cite{SenguptaChatWangYeo02}, and nonlinear incompressible Navier-Stokes equation \cite{SenguptaSumaSeng19, Sengupta20} showing the dominance of nonmodal growth.
 
Despite the spectacular growth of disturbances in a boundary layer by TCV \cite{Kendall87,Kendall90}, no clearer explanations have been proposed so far. Here, this is demonstrated including compressibility effects to show the presence of modal and nonmodal disturbances which interact among themselves to display spectacular disturbance growth by solving CNSE.

The paper is formatted in the following way. In section \ref{m-nm}, the distinction between modal and nonmodal disturbance growth is explained for the wall excitation problem. In section \ref{nm-r}, the nonmodal route of transition observed in the case of a single translating vortex in presented. The effect of a periodic, infinite train of convecting vortices (TCV) on the transition is presented in section \ref{nm-r-t} highlighting the role of nonmodal route of transition. The unsteady forcing caused by the TCV excitation is shown in section \ref{mech} and a plausible mechanism for violent turbulence encounters of an aircraft is made. The paper closes with summary and conclusions in section \ref{concl}.

\section{Distinction between modal and nonmodal disturbances for wall excitation}
\label{m-nm}
The aspects of modal and nonmodal disturbance growth for a wall excitation case is described here. Consider a flat plate excited at the wall, time harmonically at a location where the Reynolds number ($Re = U_{\infty} \delta^*/ \nu$) is 1000, based on local boundary layer displacement thickness ($\delta^*$) as the length scale and free stream speed ($U_\infty$) as the velocity scale. The time scale is chosen as $\delta^*/U_{\infty}$ for solving the governing OSE given as, 

\begin{equation}
\phi^{iv} - 2 \alpha^2 \phi^{''} + \alpha^4 \phi = i Re [ (\alpha U - \omega_0)[\phi^{''} - \alpha^2 \phi] -\alpha U^{''} \phi ] 
\end{equation}

\noindent which contains both modal and nonmodal components of the response as defined by the disturbance streamfunction in the generic space-time framework by, 
\begin{equation}
\psi_d (x,y,t) = \int_{{Br}_{\alpha}} \int_{{Br}_{\omega_0}} \phi(y,\alpha, {\omega}_0) e^{i(\alpha x - {\omega}_0 t)} d\alpha d{\omega}_0
\end{equation}
Where the symbol "Br" indicates the Bromwich contours in the complex wavenumber ($\alpha$) and circular frequency ($\omega_0$) planes following the Bromwich contour integral method (BCIM) \cite{Sengupta12, PolBrem59, Papoulis62}. For flow transition 
BCIM has been pioneered in Sengupta et al. \cite{Sengupta12, SenguptaBallNijh94}.

\begin{figure}
\centering
\includegraphics[width=1.0\linewidth]{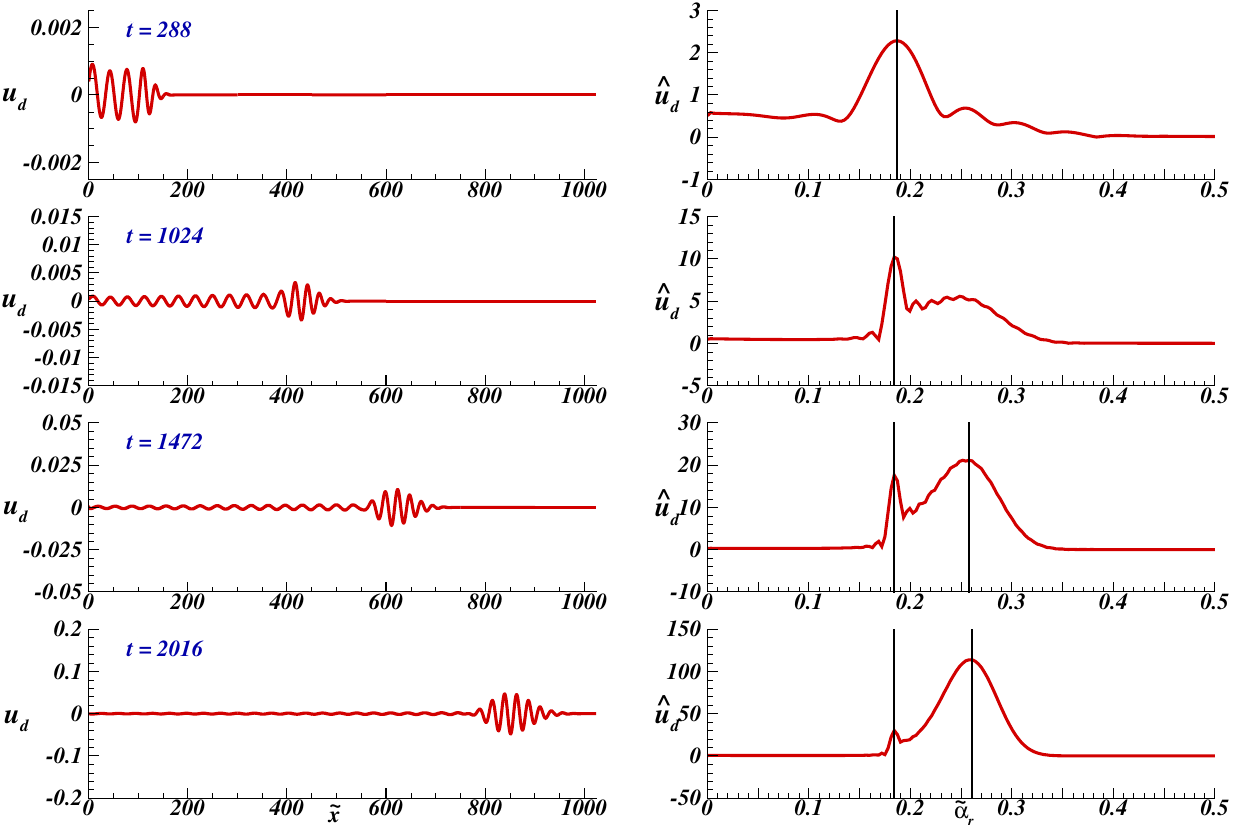}
\caption {Streamwise disturbance velocity at a height, $y  = 0.2781 \delta^*$, at the indicated times due to monochromatic frequency wall excitation for $Re=1000$ and $\omega_0 = 0.06$ shown in the left column. The corresponding spectrum are shown in the right column. The central wavenumber of TS wave ($\alpha_{TS} = 0.1840$) and STWF ($\alpha_{STWF} = 0.2608$) are indicated with lines. The indicated times are non-dimensional.}
\label{ose-ud-fft}
\end{figure}

In Fig. \ref{ose-ud-fft}, the streamwise disturbance velocity ($u_d$) is calculated from the OSE for $\omega_0=0.06$ and $Re=1000$, and shown in the left frames at the indicated times \cite{Sundaram22}. The Fourier transform of the same are shown in the right hand side frames. The spatially localized, time-harmonic exciter is placed where the Reynolds number based on the local displacement thickness is 1000. For clarity, the $y$-axis in various frames are different. In the first frame at $t=288$, one notices a distinct dominant peak, while at later times ($t= 1024$ and $1472$), one notices a second hump forming. While both these peaks keep growing with time, it is the second hump at the right that corresponds to the leading STWF dominates as it propagates downstream. The first peak always remains localized near the exciter and that has been identified as the TS wave which is the modal component with $\alpha_{TS} = 0.1840$. The STWF is the nonmodal component ($\alpha_{STWF} = 0.2608$), reported in \cite{SenguptaRaoVenk06a} for the first time to explain tsunami-like growth for a spatially stable modal component. Additionally, the local solution in the immediate vicinity of the exciter is obtained from the application of Tauber and Abel theorems, as explained in \cite{Sengupta12, PolBrem59}. It is now well established that the STWF is the main precursor of transition to turbulence by wall excitation for two- (2D) and three-dimensional (3D) flows in \cite{SenguptaBhau11, BhaumikSeng14}, respectively. Having explained the modal and nonmodal growths for harmonic wall excitation, one also notes that for this excitation only nonmodal growth is seen experimentally to be dominant, and can be obtained from the solution of OSE, as in \cite{GasterGran75, SundaramSengSeng19}, respectively. 

\section{Nonmodal route of transition for freestream excitation by a single translating vortex}
\label{nm-r}
\begin{figure}
  \centering
  \includegraphics[width=\textwidth]{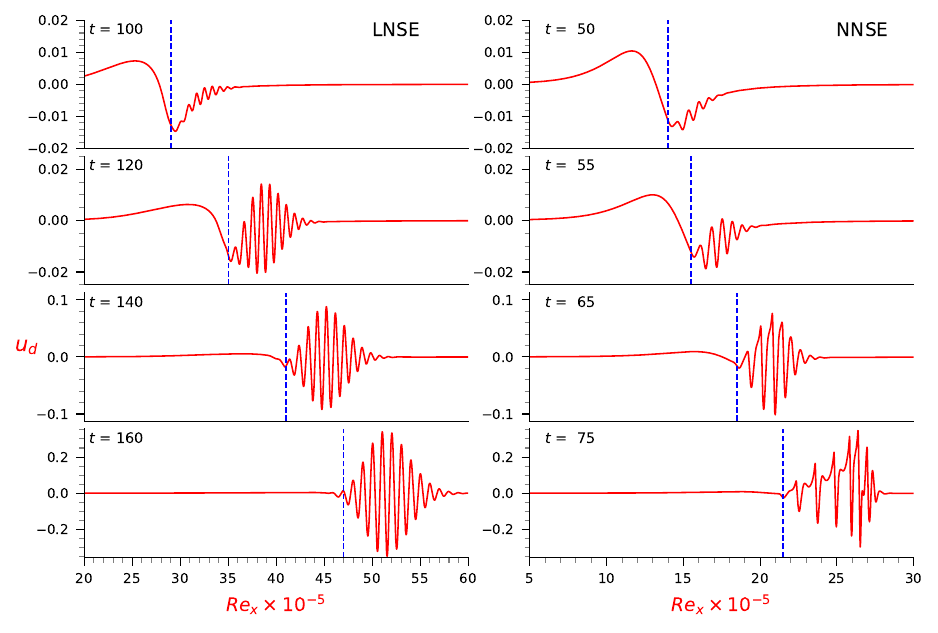}
\caption {Streamwise disturbance component $u_d$ at a height $y = 0.0028$ obtained from linearized (left) and nonlinear (right) incompressible Navier-Stokes equation simulations for a single freestream convecting vortex with parameters $\Gamma = 0.1$, $c=0.3$ and $H = 2$. Vertical dashed line indicates its instantaneous position.}
\label{Fig01}
\end{figure}

\begin{figure}
  \includegraphics[width=\textwidth]{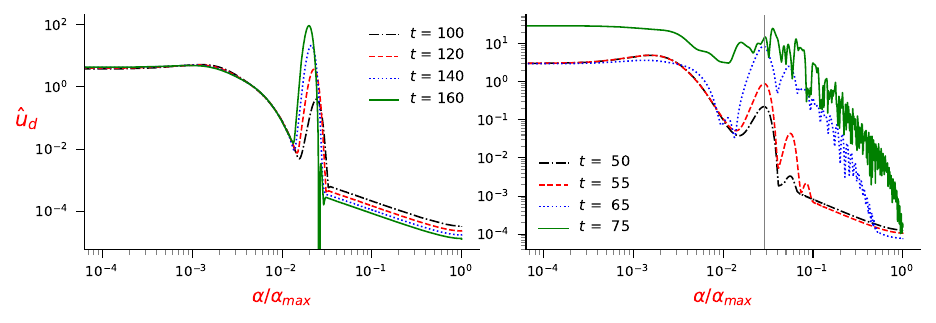}
\caption {Fourier transform of streamwise disturbance component $u_d$ at a height $y = 0.0028$ for linearized (left) and nonlinear (right) incompressible Navier-Stokes equation simulations for a single freestream convecting with parameters $\Gamma = 0.1$, $c=0.3$ and $H = 2$. The wavenumber is normalized with the maximum resolved wavenumber limit ($\alpha_{max}$).}
\label{Fig01_b}
\end{figure}

To provide a physical explanation of nonmodal growth, a typical vortex-induced disturbance growth problem is shown in Figs. \ref{Fig01} and \ref{Fig01_b}. Here, a single convecting counter-clockwise vortex (of strength $\Gamma$) in the freestream creates a response field without modal component, as noted experimentally \cite{LimSengChat04} and theoretically \cite{SenguptaSumaSeng19, Sengupta20, SenguptaSundSeng20}. Such a convecting vortex in the freestream induces disturbance growth by the action of scouring the boundary layer ahead of it for low speed of convection ($c$) or by creating an inflection point in the velocity profile for the vortex convecting at higher $c$. The latter case suffers a temporal growth following Rayleigh’s theorem \cite{Sengupta21, DrazinReid81}. The following features of the induced growth on a boundary layer by a single convecting vortex are noted: (i) There are two elements of the response, namely the local field (exactly beneath the freestream convecting vortex) and the STWF \cite{Sengupta12}; (ii) the absence of a  modal component in the response has prompted researchers to call this as the bypass transition, and (iii) for clockwise freestream vortex there is a weak interaction upstream of it. A clearer distinction between modal (TS wave) and nonmodal (STWF) component of disturbance growth by wall excitation is demonstrated in Sengupta et al. \cite{SenguptaSundSeng20}.

The global linear and nonlinear analysis \cite{Sundaram22, Sengupta20, SenguptaSundSeng20} are performed for the freestream vortex strength of $\Gamma=0.1$, convecting at $c= 0.3U_{\infty}$, and convecting at a constant height $H = 2L$, in a computational domain $(-0.5L \le x \le 120L, 0 \le y \le 1.5L)$, where L is defined from $Re_L = U_{\infty} L / \nu = 10^5$. In Fig. \ref{Fig01}, the linearized Navier-Stokes equation results are shown on the left frames, while the full nonlinear Navier-Stokes equation solutions are shown on the right. As the linear analysis has shown unlimited growth in time of the STWF \cite{SenguptaSundSeng20}, it is essential to perform a full nonlinear analysis.

In Fig. \ref{Fig01_b}, the spectrum of $u_d$  ($\hat{u}_d$) is plotted as a function of nondimensional $\alpha$ for linear (left) and nonlinear (right) incompressible Navier-Stokes equations for the case of a single freestream convecting vortex with strength $\Gamma=0.1$; convection speed $c=0.3U_{\infty}$ and for a height $H=2L$. The wavenumber is normalized in the abscissa with its maximum resolved value ($\alpha_{max}$). Due to this, result for the spectrum shown in Sengupta et al. \cite{SenguptaSundSeng20}, the nonlinear analysis demonstrates a wide-band response at higher wavenumbers, whereas a localized spectral peak is noted for the linear case.

\section{Effect of freestream periodic train of convecting vortices}
\label{nm-r-t}
\begin{figure}
    \centering
    \includegraphics[width=1.0\linewidth]{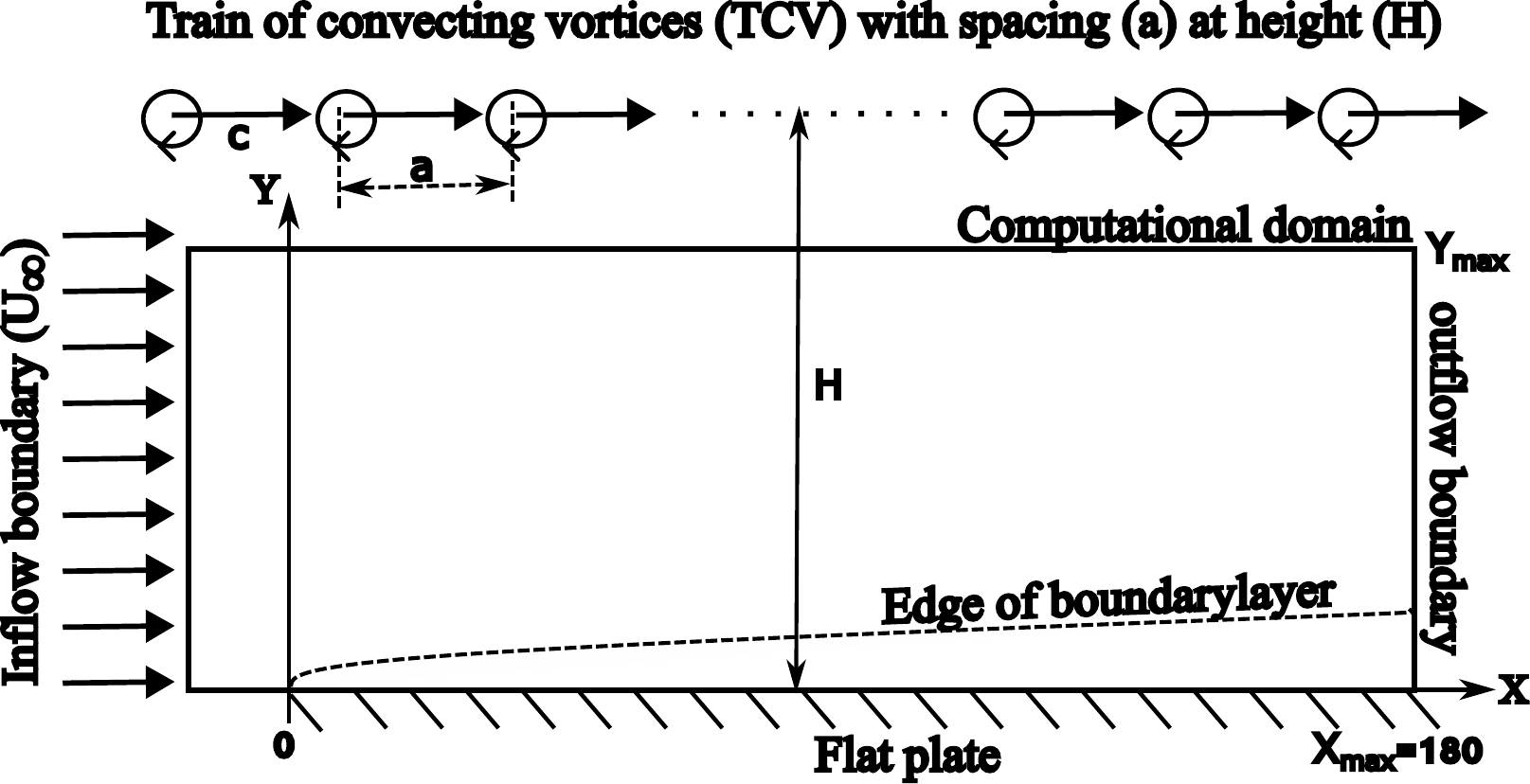}
    \caption{Schematic of the free stream excitation due to an infinite, periodic train of convecting vortices.}
    \label{Fig02}
\end{figure}

A single vortex causes moderate transient growth, shown in Fig. \ref{Fig01}. A new class of disturbance growth are caused by TCV with definitive periodicity. This is shown by computing the CNSE over a semi-infinite flat plate with the boundary layer excited by TCV of equi-spaced clockwise vortices (of strength $\Gamma =-0.005$ per unit length) at a distance $a$ apart, convecting at a constant height ($H$) and speed ($c=0.3U_\infty$) whose schematic is shown in Fig. \ref{Fig02}.
It has been shown for the case of a single convecting vortex it is the counter-clockwise vortex that shows dominant disturbance growth, whereas the clockwise vortex creates a transient growth upstream which decays very quickly resulting in non-perceptible disturbance growth. For this reason, to establish the special nature of TCV cases we have used clockwise vortices to show its uniqueness and the qualitatively different mechanism of disturbance growth. The choice of $c$ is dictated by the experimental identification of this very strong receptivity parameter in the literature \cite{Kendall87,LimSengChat04}. Secondly, a single counter-clockwise vortex only causes scouring action ahead of it to cause nonmodal growth. In the absence of any bias for the infinite TCV, we have purposely considered a clockwise vortex instead.
In Kendall \cite{Kendall87}, a relatively stronger receptivity of an equivalent TCV like experiment was reported with the maximum receptivity shown within a band of convection speed given by $0.25 \leq c/U_{\infty} \leq 0.35$. However, in the experimental setup \cite{Kendall87}, the convecting vortices were free to move in the wall normal direction and also could mutually interact for the vortices of the opposite sign thereby not showing the true potential of the TCV case. To demonstrate the physical mechanism of such encounters of TCV for the configuration shown in Fig. \ref{Fig02} was simulated to create a kernel experiment with each parameters ($\Gamma, c, H$) strictly controlled. A maximum receptivity of the TCV was noted, with the streamwise disturbance velocity achieving a value of $|u_d|\approx 0.004U_\infty$ in the experiment\cite{Kendall87} for the case of ($c=0.3U_\infty$), the 2D CNSE is used for the simulation. 
\subsection{Governing equations \& Numerical Methods}
The receptivity due to an infinite train of periodic convecting vortices can be very strong, depending upon the convection speed as has been shown by \cite{Kendall87}.
As a consequence, locally the flow can take a Mach number value that is above the critical value for which the flow can be compressible. Hence, unsteady 2D CNSE is used for the direct simulations. The nondimensional form of 2D CNSE is given next,

\begin{equation}
    \frac{\partial \hat{Q}}{\partial t}+\frac{\partial \hat{E_c}}{\partial x}+\frac{\partial \hat{F_c}}{\partial y} = \frac{\partial \hat{E_v}}{\partial x} + \frac{\partial \hat{F_v}}{\partial y}
    \label{gov_eqn}
\end{equation}

\noindent where $Q$ is the vector of conserved variables $\hat{Q} = \left[\rho ;\; \rho u;\; \rho v; \; \rho e_t\right]^{T}$; $\hat{E_c}$ and $\hat{F_c}$ are the convective fluxes $\hat{E_c} = \left[\rho u ;\; \rho u^{2} + p ;\; \rho u v ;\; \ ( \rho e_t + p) u \right]^{T}$, $\hat{F_c} = \left[\rho v ;\; \rho u v ;\; \rho v^{2} + p ;\;  \ ( \rho e_t + p) v \right]^{T}$; $\hat{E_v}$, $\hat{F_v}$ are the viscous fluxes $\hat{E_v} = \left[0;\; \tau_{xx} ;\ \tau_{xy} ;\ u \tau_{xx} + v \tau_{xy} - q_x\right]^{T}$, $\hat{F_v} = \left[0;\; \tau_{yx} ;\ \tau_{yy} ;\ u \tau_{yx} + v \tau_{yy} - q_y\right]^{T}$. $q_i$ denotes the heat flux.

The terms $\tau_{ij}$ are the non-dimensional stress tensor terms which contain the Reynolds number (Re), and are given as, 
\begin{equation}
    \tau_{xy} = \tau_{yx} = \frac{\mu}{Re} \left[\frac{\partial v}{\partial x} + \frac{\partial u}{\partial y} \right]
\end{equation}

\begin{equation}
    \tau_{xx} = \frac{1}{Re} \left (2 \mu \frac{\partial u}{\partial x} + \lambda \left[\frac{\partial u}{\partial x} + \frac{\partial v}{\partial y} \right ]\right)
\end{equation}

\begin{equation}
    \tau_{yy} = \frac{1}{Re} \left (2 \mu \frac{\partial v}{\partial y} + \lambda \left[\frac{\partial u}{\partial x} + \frac{\partial v}{\partial y} \right ]\right)
\end{equation}

\noindent Stoke's hypothesis \(\left(\lambda = -\frac{2}{3}\mu\right)\) is employed to evaluate the stress terms. 

The system of equations is closed with the perfect gas law, \(p = \rho R_{nd}T\). The nondimensional parameters involved in the CNSE are the free stream Mach number $M_\infty$, $Re_L$ as described earlier and the Prandtl number $Pr$. We have used the value of $Pr=0.72$ for the present simulations.

Non-dimensionalization is performed with proper reference scales given as,

\begin{equation}
\begin{split}
&x_{ref,i} = L ;\;\; v_{ref,i} = U_{\infty} ;\;\; t_{ref} = \frac{L}{ U_{\infty}} ;\;\; \rho_{ref} = \rho_\infty ;\;\; T_{ref}=T_\infty \\
&p_{ref} = \rho_\infty U^2_\infty ;\;\; e_{t_{ref}} = U_{\infty}^2 ;\;\; \mu_{ref} = \mu_\infty ;\;\;\; \lambda_{ref} =\lambda_\infty
\end{split}
\end{equation}

\noindent where subscript $ref$ denotes the reference quantities.

\subsection{Boundary conditions}
To determine the disturbance induced by an infinite array of irrotational freestream vortices convecting over the flat plate at inlet and top boundaries as shown in the schematic Fig. \ref{Fig02}, an image vortex system is employed. This gives rise to an induced perturbation velocity in the inviscid part of the flow. The details and expressions of these perturbation velocity components are given in the literature, as in \cite{Sengupta12, Robertson} and these expressions have been used to calculate the imposed time-dependent boundary conditions at the inflow and on the top of the computational domain, for solving the CNSE.

\subsection{Numerical Methods} 
In the present research, the CNSE is solved for higher accuracy using compact schemes for the convection terms. A sixth-order NUC6 scheme developed in Sharma et al. \cite{NUC6}, is used in the physical plane with non-uniform grid spacing. One of the major sources of accuracy is achieved due to the ability of the NUC6 scheme to function in the non-uniform grid in the physical plane itself as it removes the additional sources of aliasing error which come into play because of grid transformation. Needless to say that such a treatment also reduces aliasing in the evaluation of the nonlinear convection terms. The NUC6 scheme for a non-uniform grid as given in Sharma et al. \cite{NUC6} can be written as,

\begin{equation}
\begin{split}
	p_{j-1} \, u_{j-1}' + u_{j}' +p_{j+1} \, u_{j+1}' &= s_1 \frac{u_j - u_{j-2}}{h_{\text{llj}}}  + s_2 \frac{u_j - u_{j-1}}{h_{\text{lj}}} \\
	&+ s_3 \frac{u_{j+1} - u_j}{h_{\text{rj}}} + s_4 \frac{u_{j+2} - u_j}{h_{\text{rrj}}}
\end{split}
\end{equation}

where, $p_{j-1}$ = $W_4$  $\alpha_{j-1}$, $s_1$ = $W_4$  $q_1$ + (1-$W_4$) $r_1$, $s_2$ = $W_4$  $q_2$ + (1-$W_4$) $r_2$, $p_{j+1}$ = (1-$W_4$) $\alpha_{j+1}$, $s_3$ = $W_4$  $q_3$ + (1-$W_4$) $r_3$, $s_4$ = $W_4$  $q_4$ + (1-$W_4$) $r_4$, $W_4 = \frac{h_{rj}}{h_{rj} + h_{lj}}$ and $h_{llj}$ = ($x_j$ - $x_{j-2}$) \qquad $h_{rrj}$ = ($x_{j+2}$ - $x_{j}$), $h_{lj}$ = ($x_j$ - $x_{j-1}$) \qquad $h_{rj}$ = ($x_{j+1}$ - $x_{j}$).\\

The values of the coefficients are obtained in \cite{NUC6} as follows,

$\alpha_{j-1}$ = $\alpha_{j+1}$ = $\frac{2}{3} $

$q_1$ = $\frac{1}{6}$ \qquad $q_2$ = $\frac{11}{9}$ \qquad $q_3$ = $\frac{1}{3}$ \qquad $q_4$ = $\frac{1}{18}$

$r_1$ = - $\frac{1}{18}$ \qquad $r_2$ =  $\frac{1}{63}$ \qquad $r_3$ = $\frac{11}{9}$ \qquad $r_4$ = $\frac{1}{6}$

\subsection{Nonmodal route of transition caused by TCV}
\begin{figure}
    \centering
    \begin{subfigure} [t = 45\label{dist-vort-045-y2}] {\includegraphics[width=.95\textwidth]{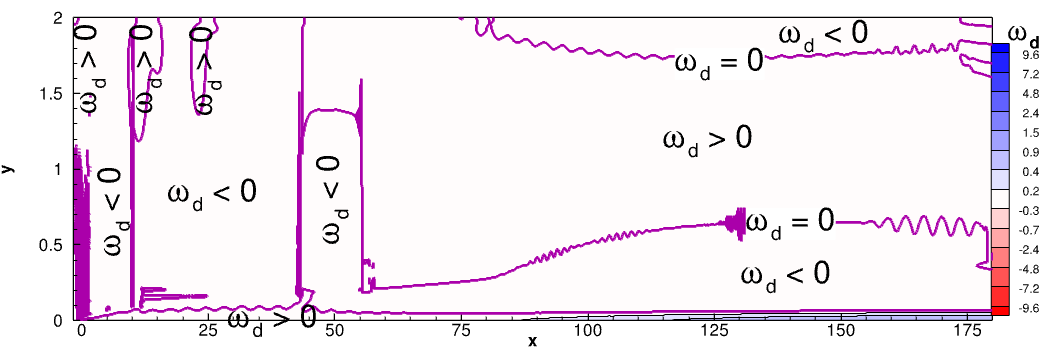}}
    \end{subfigure}
    \begin{subfigure} [Nearer to flat plate at t = 45\label{dist-vort-045}] {\includegraphics[width=.95\textwidth]{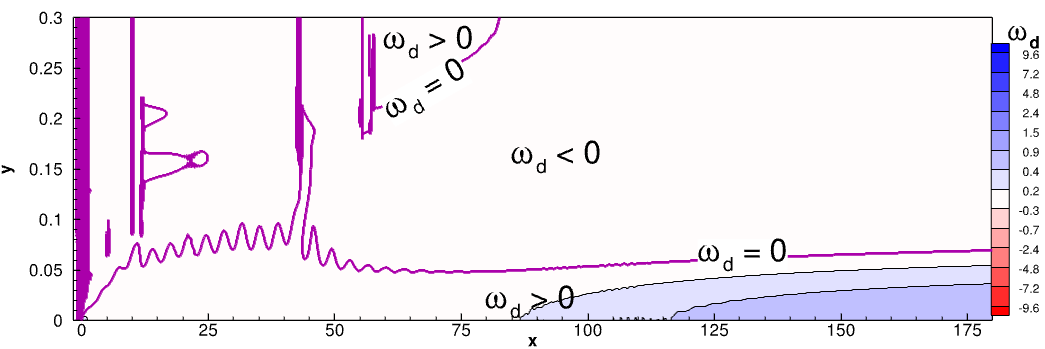}}
    \end{subfigure}
    \caption{Evolution of disturbance vorticity $\omega_d$ for a case of infinite, periodic train of convecting vortices. The convecting vortices have strength $\Gamma = -0.005$ at non-dimensional time $t=45$.}
    \label{Fig03-45}
\end{figure}

\begin{figure}
    \centering
    \begin{subfigure} [t=150\label{dist-vort-150-y2}] {\includegraphics[width=.95\textwidth]{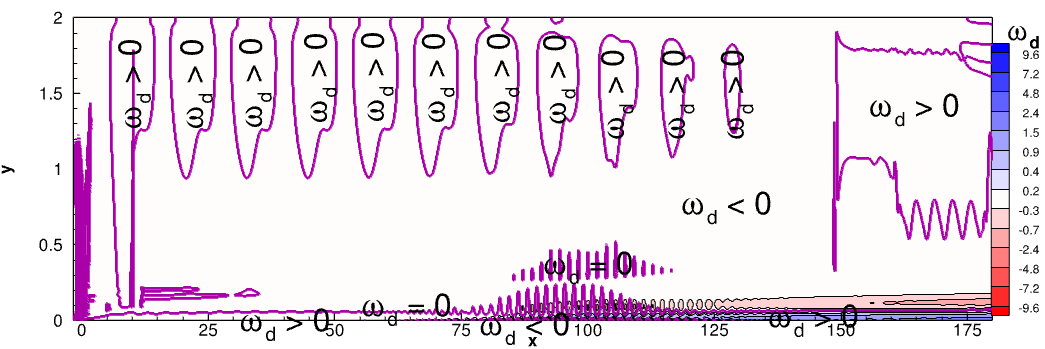}}
        \end{subfigure}
    \begin{subfigure} [Nearer to flat plate at t=150\label{dist-vort-150}] {\includegraphics[width=.95\textwidth]{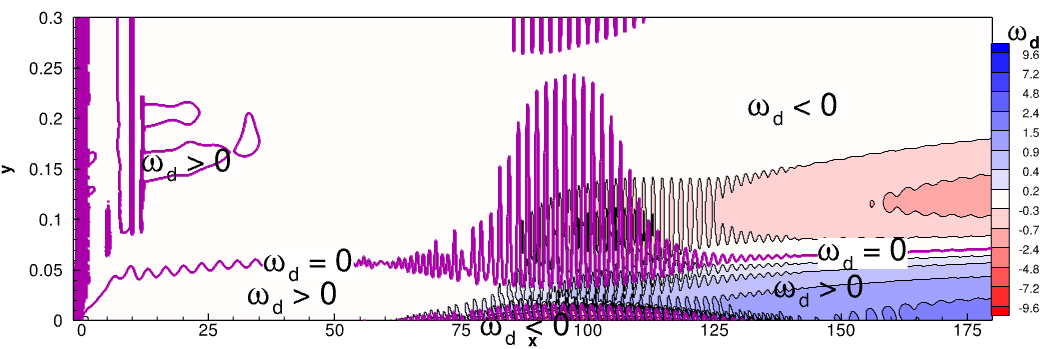}}
    \end{subfigure}
    \begin{subfigure} [Very close to flat plate at t=150\label{dist-vort-150-y0p1}] {\includegraphics[width=.95\textwidth]{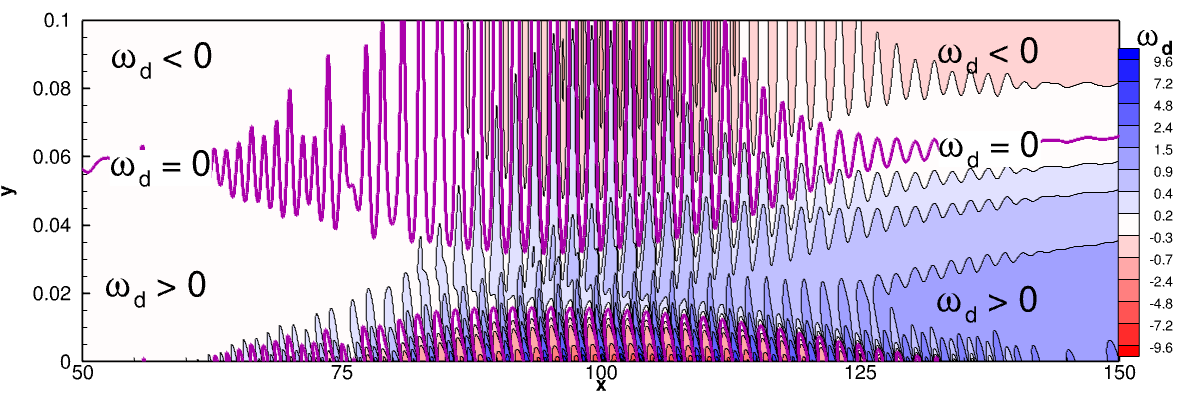}}
    \end{subfigure}
    \caption{Evolution of disturbance vorticity $\omega_d$ for a case of infinite, periodic train of convecting vortices. The convecting vortices have strength $\Gamma = -0.005$ at non-dimensional time $t=150$.}
    \label{Fig03-150}
\end{figure}

\begin{figure}
    \centering
    \begin{subfigure} [t=175\label{dist-vort-175-y2}] {\includegraphics[width=.95\textwidth]{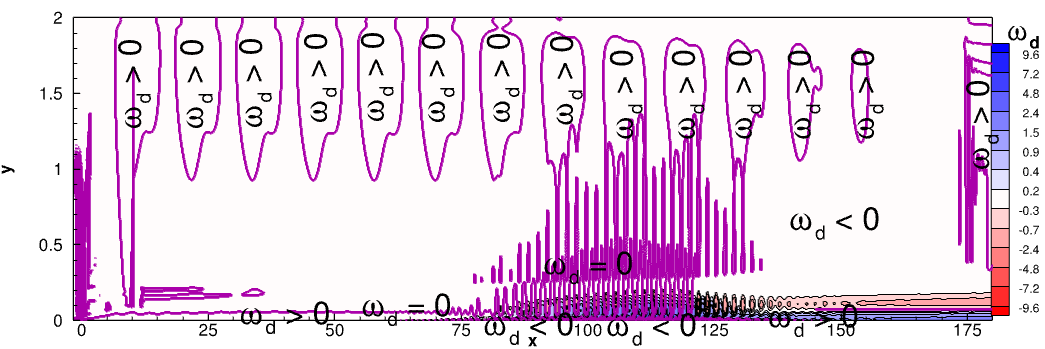}}
    \end{subfigure}
    \begin{subfigure} [Nearer to flat plate at t=175\label{dist-vort-175}] {\includegraphics[width=.95\textwidth]{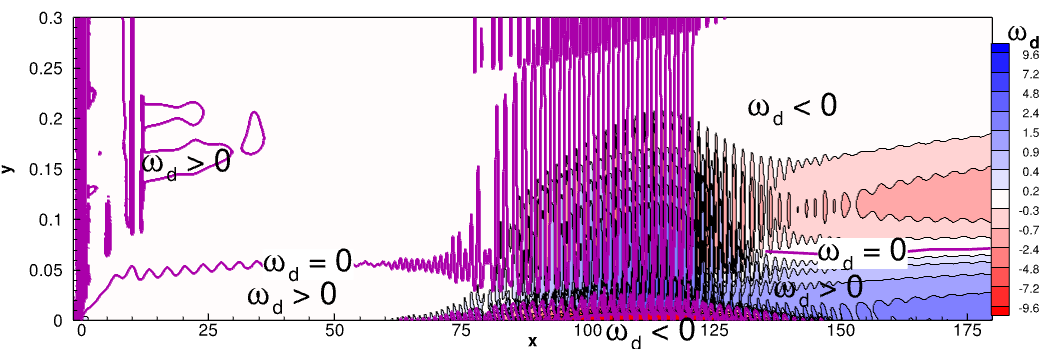}}
    \end{subfigure}
    \caption{Evolution of disturbance vorticity $\omega_d$ for a case of infinite, periodic train of convecting vortices. The convecting vortices have strength $\Gamma = -0.005$ at non-dimensional time $t=175$.}
    \label{Fig03-175}
\end{figure}

The disturbance vorticity contours of the TCV problem are shown in Figs. \ref{Fig03-45}, \ref{Fig03-150} and \ref{Fig03-175} at the indicated times in two parts: (a) in the whole computational domain and (b) in the region nearer to flat plate. In these figures, zero disturbance vorticity contour is drawn in bold (in color image: magenta) and labeled. Similarly, the regions of positive and negative disturbance vorticity are labeled accordingly along with shade in high vorticity regions. 
At $t=0$, the TCV disturbance is impulsively imposed. 
For the earliest shown time ($t=45$), very weak disturbances are seen in the major part of the domain as shown in Fig. \ref{dist-vort-045-y2}. Towards the inflow boundary there are two smaller loops (except the loop present adjacent to inflow boundary) of positive $\omega_d$ adjacent to top edge due to TCV. By considering these loops there is triple deck like structure of $\omega_d$ with positive region on flat plate, negative away from it and once again positive in loops. But, away from inflow boundary in streamwise direction triple deck converts into quadruple deck with a negative zone adjacent to top edge. Upon zooming near to flat plate, Fig. \ref{dist-vort-045} reveals that there is higher positive $\omega_d$ region (shaded) from mid of the domain to exit, which is very much embedded inside the boundary layer.

At $t=150$, the imprint of TCV on top edge progresses down-stream with eleven loops of positive $\omega_d$ while reducing the extent of negative $\omega_d$ region and thus quadruple-deck region as depicted in Fig. \ref{dist-vort-150-y2}. In Fig. \ref{dist-vort-150} it is seen that the region with higher $\omega_d$ (shaded) extends in both streamwise and wall-normal directions. At this time, one notices a little region of alternative positive and negative $\omega_d$ in saw tooth form (called as saw tooth deck) over flat plate around $x=100$, where flow transition to turbulence took place as shown in Fig. \ref{dist-vort-150-y0p1}. In wall-normal direction vortical disturbances are stretched as vortical eruptions spreading across different decks of $\omega_d$. Locally the triple-deck structure became 
quintuple-deck like
structure around $x=100$ with two saw tooth decks: one on wall and one more isolated deck at around $y=0.5$.
Such vortical eruptions are noted to become more prominent with increased extent in wall-normal direction at $t=175$ as shown in Fig. \ref{dist-vort-175-y2}. At $t=175$, the saw tooth isolated deck of $t=150$  (i) erupts upwards till the positive $\omega_d$ loops present adjacent to the top edge and (ii) erupts downwards till the edge of negative $\omega_d$ region with interlocking manner overlap (iii) besides extending in both upstream and downstream directions.

For this TCV case, both modal and nonmodal components are present in the spectrum which interact due to nonlinear dynamics, as shown in Figs. \ref{Fig04} and \ref{Fig05} showing the streamwise and wall-normal disturbance velocity components as a function of $x$ in the left columns. The corresponding Fourier transform are shown in the right columns. 

Corresponding to the spacing ($a$), the modal components are expected to be present as multiples of $\alpha_0\; (=2\pi/a)$ in the spectrum. These modal components are shown by red vertical lines in the spectrum. 
The additional nonmodal structures in the spectrum are marked as $N_1$, $N_2$, ..... etc. 
The location and their appearances are decided by the nonlinear dynamics of the NSE. 
Note the vertical scale in the bottom frame is zoomed out, indicating the rapidity with which the disturbance field grows at later times. It is noted that the spectral peaks and amplitudes increase with time from the $y$-scales used in the plot. At $t=150$, the spectral amplitudes of $u_d$ are similar for the two heights whereas at $t=175$, the amplitudes are significantly higher for the lower height. 

For the wall-normal disturbance velocity component $v_d$, a complementary nature is observed in Fig. \ref{Fig05}. This component of velocity is strongly inhibited near the wall, whereas it is prominent near the freestream at $t=150$ and $175$. At $t=150$, the modal peaks corresponding to $2\alpha_0$, $4\alpha_0$ and $6\alpha_0$ are dominant whereas by $t=175$, the effect of nonlinearity is clearly seen as the nonmodal components are dominant.

\section{Unsteady forcing during TCV effects}
\label{mech}
\begin{figure}
\centering
\includegraphics[width=\textwidth]{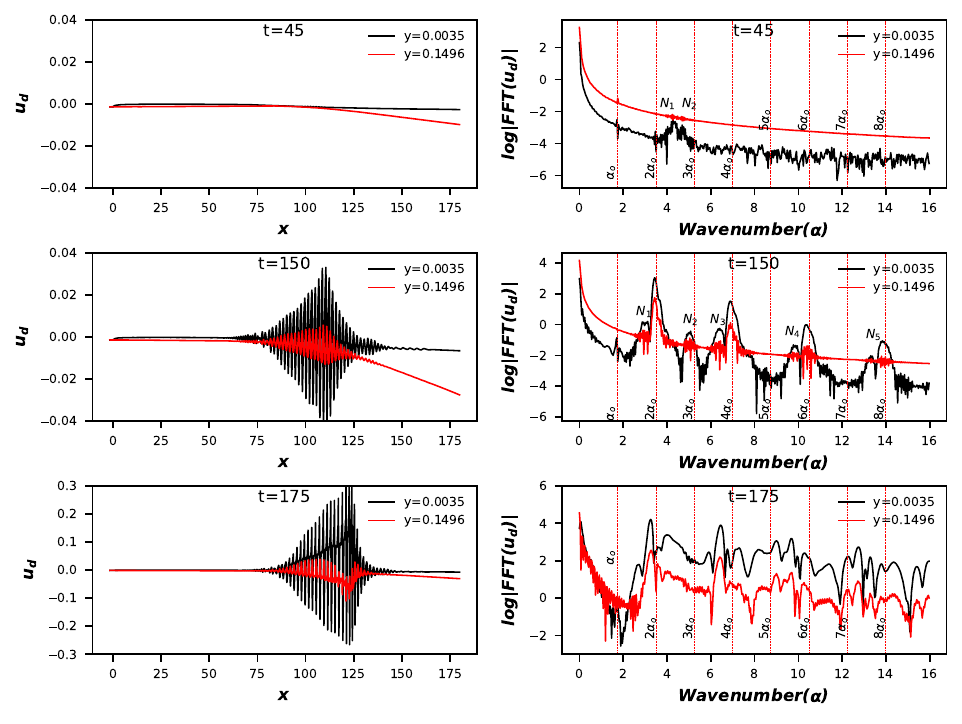}
\caption{Streamwise disturbance component $u_d$ (left) and its Fourier transform (right) at two indicated heights for the free stream excitation by train of infinite vortices of strength $\Gamma = -0.005$. The modal peaks $\alpha$ to $8\alpha$ are fixed by the spacing between the vortices and the nonmodal peaks are marked as $N_j$.}
\label{Fig04}
\end{figure}

\begin{figure}
\centering
\includegraphics[width=\textwidth]{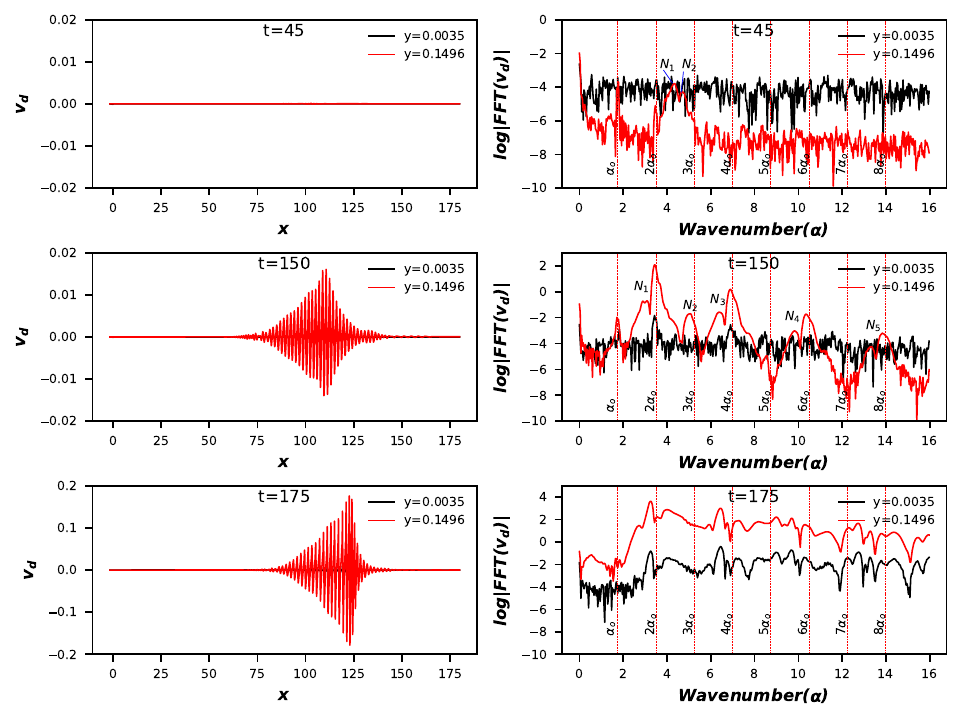}
\caption{Wall-normal disturbance component $v_d$ (left) and its Fourier transform (right) at two indicated heights for the free stream excitation by train of infinite vortices of strength $\Gamma = -0.005$. The modal peaks $\alpha$ to $8\alpha$ are fixed by the spacing between the vortices and the nonmodal peaks are marked as $N_j$.}
\label{Fig05}
\end{figure}

\begin{figure}
\centering
\includegraphics[width=1.0\linewidth]{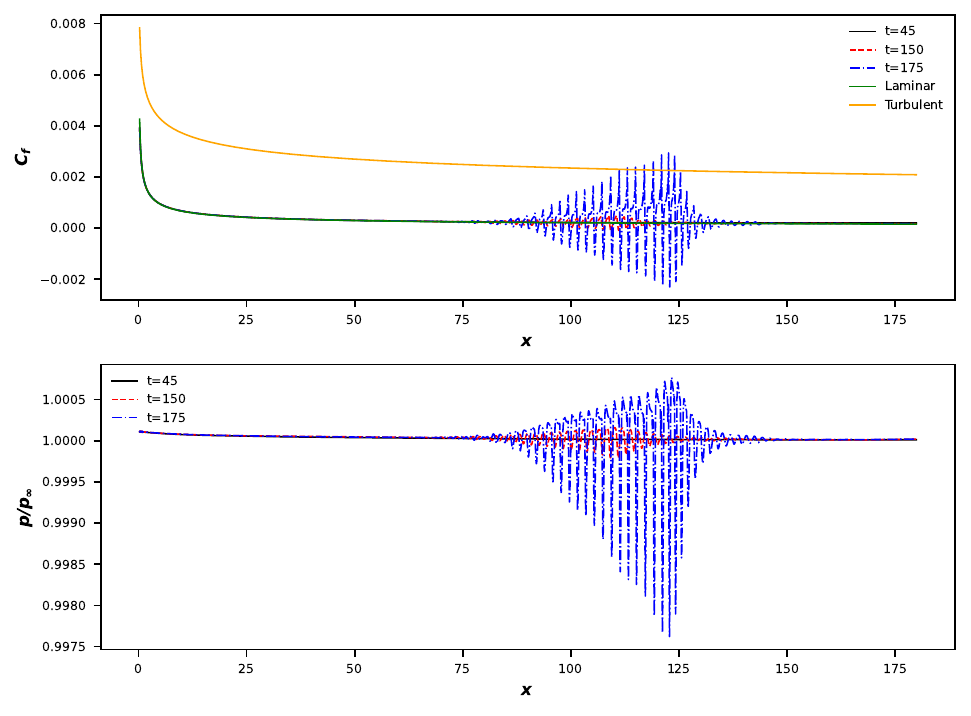}
\caption{Evolution of skin-friction coefficient $C_f$ and pressure on the surface for the free stream excitation by train of infinite vortices (clockwise of strength $\Gamma = -0.005$), respectively.}
\label{Fig06}
\end{figure}

From Figs. \ref{Fig03-45} to \ref{Fig05}, we have seen distinctly unsteady behaviour of the velocity and vorticity fields. Thus it is expected that there would be significant unsteady forcing caused by the TCV. In Fig. \ref{Fig06}, we have shown the wall skin friction and the surface pressure on the flat plate at the indicated times. For the sake of reference, typical skin friction variations over a flat plate are shown for the laminar and fully developed turbulent flow (time averaged). It is clearly noted that both these integrated quantities are increasing with time in magnitude. Furthermore, it is noted that the maximum instantaneous wall skin friction increases above the turbulent value. One can also see that these two aerodynamic quantities contribute to the lift and drag experienced by the plate. The increased suction indicated by the pressure is consistent with $v_d$ shown in Fig. \ref{Fig05}.

As the unsteady forcing is localized in TCV actions, these will similarly give rise to unsettling aerodynamic moments affecting the flight dynamics of an equivalent aircraft wing under the action of TCV. Similar effects can also be expected to arise for the flow field over the fuselage and other aerodynamic empennages caused by TCV. 

\section{Summary and Conclusions}
\label{concl}
In summarizing the research reported here, we need to distinguish between modal and nonmodal disturbance field clearly, as it has been done here with the help of BCIM of Orr-Sommerfeld equation (in Fig. \ref{ose-ud-fft}) showing the presence of modal, nonmodal and local solution component for transition caused by wall excitation. However, to explain the turbulence encounters of aircraft,
it is important to appreciate that such effects are related to vortical excitation from the free stream, for which two canonical problems are studied: (i) An aperiodic translating vortex convecting over a flat plate (that mimics the suction surface of an aircraft wing) studied experimentally \cite{LimSengChat04} and theoretically \cite{SenguptaSumaSeng19, SenguptaSundSeng20, Sengupta12,Sengupta21}. Such free stream excitation does not display modal component (for the bypass transition), and is effective only for counter-clockwise (anticyclonic) vortex to display mild to moderate transient response field, referred to 
as CAT in the literature \cite{SenguptaLimChat02, DuttonPano70, ClarkHallKerr00,
WingroveBach94, BrowningWatkStar70} for diagnostic and conjectural models. Here, the physical explanation is provided (in Fig. 2) for CAT-like moderate interactions, which happens for a narrow speed range of the translating counter clockwise vortex (chosen here as $c= 0.3U_{\infty}$). In this figure, the global linear mechanism obtained by solving the linearized NSE is compared with the solution obtained by solving the full CNSE to show the relevance of the latter over the former. In Fig. 3, the Fourier-Laplace transforms of the linear and nonlinear solutions are compared to show the correctness of the nonlinear, nonmodal route, as the linear route can provide the correct onset, but the growth of the STWF is unbounded for the linearized case, which is physically infeasible. (ii) The TCV problem is more generic for effects of free stream turbulence, as compared to the case of disturbance field created by an isolated convecting vortex in the free stream. Thus, for the schematic shown in Fig. 4, stronger interaction can occur, when a modal length scale in the input spectrum is introduced by the TCV. Here, interactions can occur for any sign of the constituents of TCVs. The corresponding vortical encounters by TCV are much stronger. Apart from the nonmodal component, the modal components are noted which also display vigorous interactions in the spectrum, even when the strength of the vortices are many orders of magnitude lower. Typical results are shown in Figs. 5 to 7, where a rapid growth of the disturbance is noted with a continuous spectrum typical of nonmodal components, starting at very early times.

\section{Acknowledgments}
\label{ack}
The early contributions of Mr. Vipin Sharma, Graduate student, Department of Mechanical Engineering, IIT (ISM) Dhanbad are highly acknowledged.


%
%

%


\bibliography{TCV_References_mod}

\end{document}